\documentclass[aps,pra,reprint,showpacs]{revtex4-1}
\usepackage[latin1]{inputenc}
\usepackage{amsmath}
\usepackage{amsfonts}
\usepackage{amssymb}
\usepackage{amsthm}
\usepackage{bm}
\usepackage{mathrsfs}
\usepackage{srcltx}
\usepackage{srctex}
\usepackage{tensor}

\begin{document}

\title{Transformation Optics in Nonvacuum Initial Dielectric Media}
\author{Robert T. Thompson}
\email{robert@cosmos.phy.tufts.edu}
\affiliation{Department of Mathematics and Statistics,
University of Otago,
P.O. Box 56, 
Dunedin, 9054,  New Zealand
}

\begin{abstract}
Previous formulations of transformation optics have generally been restricted to transformations from relatively simple initial media, such as the vacuum, because of limitations due to their non-covariance.  I show that a completely covariant approach enables arbitrary transformations from arbitrarily complex initial linear dielectrics. 
\end{abstract}

\pacs{42.15.Eq, 41.20.Jb, 02.40.Hw, 42.25.-p}

\maketitle
\section{Introduction}
The nascent field of metamaterial based transformation optics has garnered much attention in the past few years, beginning with the highly publicized development of an electromagnetic cloak \cite{Pendry:2006a,Leonhardt:2006a,Schurig:2006,Rahm:200887}. The original approach to transformation optics was based on purely spatial coordinate transformations \cite{Pendry:2006a,Schurig:2006,Milton:2006}.  A step toward generalizing the allowed transformations \cite{Leonhardt:2006ai} was based on a similarity between the constitutive equations for electromagnetic fields in a curved vacuum space-time to the constitutive equations for electromagnetic fields in materials residing in Minkowski space-time developed by Plebanski and De Felice \cite{Plebanski:1959ff,DeFelice:1971}.  However, the Plebanski-De Felice equations are not strictly covariant, as cautioned by Plebanski himself.  Furthermore, the use of the Plebanski-De Felice equations in transformation optics is somewhat unsatisfying because it requires identifying a coordinate transformation in flat space-time with a curved manifold in a rather nonrigorous manner.  More recently a completely covariant and manifestly four-dimensional approach to transformation optics has been developed that is both more rigorous and more widely applicable \cite{Thompson:2010a,Thompson:2010b}.

It seems likely that many useful technological applications of transformation optics will require, as a design specification, that the apparatus operate in a nonvacuum environment, such as water or a coolant fluid.  Such a scenario has mostly fallen outside the domain of validity of transformation optics because the Plebanski-De Felice equations are strictly only valid for vacuum space-times.  It might be expected that, for isotropic prior dielectrics, the constant parameters of the prior simply replace the vacuum values \cite{Leonhardt:2006ai}, which turns out to be true in that special case, but in general it is not possible to treat arbitrary prior materials with the Plebanski-De Felice equations because the vacuum is always isotropic and nonmagnetoelectric whereas dielectric media are not.  At the level of Maxwell's equations in three dimensions, other approaches \cite{Pendry:2006a,Milton:2006,Cheng:2009} have the potential to accommodate prior dielectrics with matrix-valued permeability and permittivity for special cases of purely spatial transformations.  Those approaches are somewhat limited in scope by failing to incorporate the most general transformations, which include time \cite{Cummer:2010b}, and do not include more complicated material parameters such as magnetoelectric couplings.

I demonstrate that a completely covariant and manifestly four-dimensional approach to transformation optics allows for general transformations within arbitrary prior nonvacuum material distributions.  While purely spatial transformations from relatively simple initial materials agree with what might be expected from other approaches, certain classes of transformations and more complicated materials require the completely covariant approach, where I find that the resulting material parameters depend on the initial material parameters in complicated and unexpected ways.

This paper is organized as follows.  Section \ref{Sec:EandM} briefly  summarizes the completely covariant, manifestly four-dimensional description of classical electrodynamics.  Section \ref{Sec:TO} sketches the completely covariant approach to transformation optics developed in Refs.\ \cite{Thompson:2010a,Thompson:2010b} and presents the main result found therein. Sections \ref{Sec:Isotropic} -- \ref{Sec:MagnetoElectric} examine a variety of transformations within various prior dielectric media, including isotropic and anisotropic media with and without magnetoelectric couplings.  To highlight the effects of the prior dielectric media the transformations are kept quite general, except for the example of a square cloak embedded in an anisotropic, nonmagnetoelectric media, shown in Sec.\ \ref{Sec:SquareCloak}.  I conclude with some discussion in Sec.\ \ref{Sec:Conclusions}.

\section{Classical Electrodynamics} \label{Sec:EandM}
Assume space-time to consist of a manifold $M$ and metric $\mathbf{g}$. The electric field $\vec{E}$ and magnetic flux $\vec{B}$ are combined into a single mathematical object, the covariant field strength tensor $\mathbf{F}$, that in a local Cartesian frame or Minkowski space-time has the matrix representation
\begin{equation} \label{Eq:FComponents}
 F_{\mu\nu} = \left(
 \begin{matrix}
  0 & -E_x & -E_y & -E_z\\
  E_x & 0 & B_z & -B_y\\
  E_y & -B_z & 0 & B_x\\
  E_z & B_y & -B_x & 0
 \end{matrix}
 \right).
\end{equation}
Additionally, the electric flux $\vec{D}$ and magnetic field $\vec{H}$ are combined into the covariant excitation tensor $\mathbf{G}$, that in a local Cartesian frame or Minkowski space-time has the matrix representation
\begin{equation} \label{Eq:GComponents}
 G_{\mu\nu} = \left(
 \begin{matrix}
  0 & H_x & H_y & H_z\\
  -H_x & 0 & D_z & -D_y\\
  -H_y & -D_z & 0 & D_x\\
  -H_z & D_y & -D_x & 0
 \end{matrix}
 \right).
\end{equation}
Maxwell's equations are succinctly expressed as
$\mathrm{d}\mathbf{F}=0$, and $\mathrm{d}\mathbf{G}=\mathbf{J}$,
where $\mathrm{d}$ is the exterior derivative, and $\mathbf{J}$ is the charge-current 3-form (see, e.g., Ref.\ \cite{Misner:1974qy}).  

Furthermore, in a linear dielectric medium there exists a relationship between $\mathbf{F}$ and $\mathbf{G}$ given by the constitutive equation \cite{Thompson:2010a,Thompson:2010b}
\begin{equation} \label{Eq:Constitutive}
 \mathbf{G} = \boldsymbol{\chi}(\star\mathbf{F}),
\end{equation}
which in component form reads
\begin{equation} \label{Eq:ConstitutiveIndices}
 G_{\mu\nu} = \chi\indices{_{\mu\nu}^{\alpha\beta}}\star\indices{_{\alpha\beta}^{\sigma\rho}}F_{\sigma\rho}.
\end{equation}
In Eq.\ (\ref{Eq:Constitutive}), $\star$ is the Hodge dual on $(M,\mathbf{g})$, which for present purposes is to be understood as a map from 2-forms to 2-forms that has components
\begin{equation} \label{Eq:star}
 \star\indices{_{\alpha\beta}^{\mu\nu}} = \frac12 \sqrt{|g|}\epsilon_{\alpha\beta\sigma\rho}g^{\sigma\mu}g^{\rho\nu}.
\end{equation}
The tensor $\boldsymbol{\chi}$ contains information on the dielectric material's properties such as permittivity, permeability, and magnetoelectric couplings.  We take $\boldsymbol{\chi}$ to be independently antisymmetric on its first two and last two indices, and in vacuum $\boldsymbol{\chi}(\star\mathbf{F}) = \star\mathbf{F}$.  This last condition means that the classical vacuum is treated as a linear dielectric with trivial $\boldsymbol{\chi}$, recovering the usual, trivial, constitutive relations in vacuum.

The components of the constitutive equation provide a set of six independent equations that in Minkowski space-time can be collected in the form
\begin{equation} \label{Eq:ConstitutiveComponents1}
 H_a=(\check{\mu}^{-1})\indices{_a^b}B_b + (\check{\gamma_1}^*)\indices{_a^b}E_b, \  D_a=(\check{\varepsilon}^*)\indices{_a^b}E_b+ (\check{\gamma_2}^*)\indices{_a^b}B_b,
\end{equation}
where the notation $\check{a}$ denotes a $3\times 3$ matrix.  Rearranging these to 
\begin{equation} \label{Eq:ConstitutiveComponents2}
 B_a=(\check{\mu})\indices{_a^b}H_b + (\check{\gamma_1})\indices{_a^b}E_b, \  D_a=(\check{\varepsilon})\indices{_a^b}E_b+ (\check{\gamma_2})\indices{_a^b}H_b,
\end{equation}
gives a representation that may be more familiar and in which subsequent results will be expressed.  These three-dimensional representations of the completely covariant Eq.\ (\ref{Eq:Constitutive}) are essentially equivalent, and it is a simple matter to switch between them using the relations
\begin{equation} \label{Eq:ConstitutiveShift}
 \check{\varepsilon}=\check{\varepsilon}^*-\check{\gamma_2}^*\check{\mu}\check{\gamma_1}^*, \  \check{\gamma_1}=\text{-}\check{\mu}\check{\gamma_1}^*, \  \check{\gamma_2} = \check{\gamma_2}^*\check{\mu}.
\end{equation}
However, one should be aware that these $3\times 3$ matrices are not strictly tensors but simply components of $\boldsymbol{\chi}$ that have been collected into matrices.  

\section{Transformation Optics} \label{Sec:TO}
To understand transformation optics, start with an initial space-time manifold $(M,\mathbf{g},\star)$, field configuration $(\mathbf{F},\mathbf{G},\mathbf{J})$, and material distribution $\boldsymbol{\chi}$, where $\mathrm{d}\mathbf{F}=0$, $\mathrm{d}\mathbf{G}=\mathbf{J}$, and $\mathbf{G}=\boldsymbol{\chi}(\star\mathbf{F})$.  Imagine now a map $T:M\to \tilde{M}\subseteq M$ that maps $M$ to some image $\tilde{M}$ and transforms the electromagnetic fields in some smooth way to a new configuration $(\tilde{\mathbf{F}},\tilde{\mathbf{G}},\tilde{\mathbf{J}})$.  Because the underlying space-time is physically unaltered the manifold is still described by $(M,\mathbf{g},\star)$.  But for the new field configuration to be physically supported, there must exist a new material distribution $\tilde{\boldsymbol{\chi}}$.  Therefore $\mathrm{d}\tilde{\mathbf{F}}=0$, $\mathrm{d}\tilde{\mathbf{G}}=\tilde{\mathbf{J}}$, and $\tilde{\mathbf{G}}= \tilde{\boldsymbol{\chi}}(\star\tilde{\mathbf{F}})$ holds on $\tilde{M}$.  Such a transformation could, for example, map $M$ to an image $\tilde{M}$ that contains a hole, i.e.\ a region from which the fields will be excluded, as in the case of an electromagnetic cloak.   

Using the inverse, $\mathcal{T}$, of the map $T$, the initial $\mathbf{F}$ and $\mathbf{G}$ are related to the final $\tilde{\mathbf{F}}$ and $\tilde{\mathbf{G}}$ by the pullback of $\mathcal{T}$, denoted as $\mathcal{T}^*$.  This implies
\begin{equation}
 \tilde{\mathbf{G}} = \mathcal{T}^*(\mathbf{G}) = \mathcal{T}^*(\boldsymbol{\chi}(\star\mathbf{F})) = \tilde{\boldsymbol{\chi}}(\star\mathcal{T}^*(\mathbf{F})),
\end{equation}
which can be solved for $\tilde{\boldsymbol{\chi}}$ as a function of $x\in\tilde{M}$, giving \cite{Thompson:2010a,Thompson:2010b}
\begin{multline} \label{Eq:MaterialChi}
 \tilde{\chi}\indices{_{\lambda\kappa}^{\tau\eta}}(x)=\\ -\Lambda\indices{^{\alpha}_{\lambda}} \Lambda\indices{^{\beta}_{\kappa}} \chi\indices{_{\alpha\beta}^{\mu\nu}}\Big|_{\mathcal{T}(x)} \star\indices{_{\mu\nu}^{\sigma\rho}} (\Lambda^{-1})\indices{^{\pi}_{\sigma}}(\Lambda^{-1})\indices{^{\theta}_{\rho}}\, \star\indices{_{\pi\theta}^{\tau\eta}}.
\end{multline}
In Eq.\ (\ref{Eq:MaterialChi}) $\boldsymbol{\Lambda}$ is the Jacobian matrix of $\mathcal{T}$, $\boldsymbol{\Lambda}^{-1}$ is the matrix inverse of $\boldsymbol{\Lambda}$,  and in solving for $\tilde{\boldsymbol{\chi}}$ we have made use of the fact that on a four-dimensional Lorentzian manifold, acting twice with $\star$ returns the negative, $\star\star\mathbf{F}=-\mathbf{F}$.
Note that the initial material tensor $\boldsymbol{\chi}$ must be evaluated at $\mathcal{T}(x)$, while everything else is evaluated at $x$.  Equation (\ref{Eq:MaterialChi}) represents the core of transformation optics in linear dielectric materials.  Because $\chi\indices{_{\alpha\beta}^{\mu\nu}}$ on the right hand side of Eq.\ (\ref{Eq:MaterialChi}) need not be for vacuum, we can use this to examine transformation optics in nonvacuum initial, or prior, dielectric media.

\section{Isotropic Prior Dielectric} \label{Sec:Isotropic}
Suppose we wish to impose a transformation within a prior nonvacuum dielectric material residing in Minkowski space-time, such that the final result will be a dielectric material embedded in the same material.  Assume Cartesian coordinates and let the prior dielectric be at rest with respect to the laboratory system in which the electromagnetic fields are measured.  Begin, for simplicity, by letting the prior dielectric be isotropic with parameters $\mu^p$ and $\varepsilon^p$, and with vanishing magnetoelectric coupling. 

\subsection{Spatial and Temporal Scaling}
Consider the transformation
\begin{equation} \label{Eq:YInversion}
 T(t',x',y',z')=(t,x,y,z) = (t',x',\text{-}ay',z')
\end{equation}
applied in some region $y_1\leq y \leq y_2$.  This spatial coordinate scaling has been previously studied and forms the basis of a so-called superlens \cite{PhysRevLett.85.3966,Leonhardt:2006a}.  If the transformation takes place in vacuum Minkowski space-time, the well known results for the corresponding material parameters are $\check{\varepsilon}=\check{\mu} = \text{-}diag(a^{\text{-}1},a,a^{\text{-}1})$, and $\check{\gamma_1}=\check{\gamma_2}=0$, in the representation of Eq.\ (\ref{Eq:ConstitutiveComponents2}).

Allowing $\boldsymbol{\chi}$ in Eq.\ (\ref{Eq:MaterialChi}) to describe an isotropic material rather than vacuum and calculating $\tilde{\boldsymbol{\chi}}$, one instead finds parameters $\check{\varepsilon} = \text{-}\varepsilon^p diag(a^{\text{-}1},a,a^{\text{-}1})$, $\check{\mu} = \text{-}\mu^p diag(a^{\text{-}1},a,a^{\text{-}1})$, and $\check{\gamma_1}=\check{\gamma_2}=0$.   This simple result is perhaps not so surprising, and might be guessed in advance by simply replacing the vacuum parameters with the material parameters.

Next, consider a similar scaling transformation applied to the time component
\begin{equation} \label{Eq:TInversion}
 T(t',x',y',z')=(t,x,y,z) = (\text{-}at',x',y',z')
\end{equation}
during some time interval $t_1\leq t \leq t_2$.  It may be readily shown \cite{Thompson:2010a} that such a transformation in vacuum corresponds to an isotropic material $\check{\varepsilon}=\check{\mu}=\text{-}a$, with vanishing magnetoelectric coupling.  Repeating the calculation with a prior isotropic dielectric, Eq.\ (\ref{Eq:MaterialChi}) returns the not unexpected result $\check{\varepsilon}= \text{-}a\varepsilon^p$, $\check{\mu}=\text{-}a\mu^p$, and $\check{\gamma_1}=\check{\gamma_2}=0$.

\subsection{Spatially Dependent Time Transformation}
Until now it has been, for the sake of illustration, assumed that $\mathcal{T}=T^{-1}$.  But there is no guarantee that $T^{-1}$ exists.  It is apparent from Eq.\ (\ref{Eq:MaterialChi}) that the map of interest is actually $\mathcal{T}$ rather than $T$.  Henceforth, to avoid any issues related to the existence of $T^{-1}$, we refer only to a map from $\tilde{M}$ to $M$ which, for the sake of consistent notation, we continue to refer to as $\mathcal{T}$ and assume it is well-defined on $\tilde{M}$. 

Consider the slightly more non-trivial transformation
\begin{equation} \label{Eq:TimeTrans}
 \mathcal{T}(t,x,y,z)=(t',x',y',z')=(f(x)t,x,y,z)
\end{equation}
which mixes space and time with an arbitrary function $f(x)$. A specific example of this type of transformation has been previously considered in more detail for an initial vacuum Minkowski space-time \cite{Cummer:2010b}.
Turning the crank on Eq.\ (\ref{Eq:MaterialChi}) with the prior isotropic dielectric $\boldsymbol{\chi}$, extracting the material parameters from $\tilde{\boldsymbol{\chi}}$, and using Eq.\ (\ref{Eq:ConstitutiveShift}) to express the results in the representation of Eq.\ (\ref{Eq:ConstitutiveComponents2}) returns
\begin{equation} \label{Eq:IsoSTtrans}
\begin{gathered}
 \check{\varepsilon} = \frac{\varepsilon^p}{f(x)} 
 \begin{pmatrix}
  1&0&0\\
  0&1&0\\
  0&0&1
 \end{pmatrix}, 
\ 
 \check{\mu}= \frac{\mu^p }{f(x)}
 \begin{pmatrix}
  1&0&0\\
  0&1&0\\
  0&0&1
 \end{pmatrix},\\
\check{\gamma_1} = \check{\gamma_2}^{\mathtt{T}} = \frac{f'(x)t}{f(x)}
 \begin{pmatrix}
  0 & 0 & 0 \\
  0 & 0 & 1 \\
  0 & \text{-}1 & 0
 \end{pmatrix},
\end{gathered}
\end{equation}

which is again scaled by the material parameters just as before.  So far, none of these results appear to be particularly unusual or unexpected, but consider the next example.

\subsection{Time Dependent Spatial Transformation}
In Ref.\ \cite{Thompson:2010a} it was observed that magnetoelectric coupling terms seem to arise whenever space and time are mixed in either a spatially dependent time transformation or a time dependent spatial transformation.  It is easy to see that the magnetoelectric couplings of Eq.\ (\ref{Eq:IsoSTtrans}) come from the spatial dependent time transformation.  It is then natural to next consider the transformation
\begin{equation} \label{Eq:SpaceTrans}
 \mathcal{T}(t,x,y,z)=(t,x,g(t)y,z),
\end{equation}
which mixes space and time with an arbitrary function $g(t)$.  Turning the crank on Eq.\ (\ref{Eq:MaterialChi}) with a prior isotropic dielectric returns the material parameters
\begin{equation} \label{Eq:IsoTStrans}
\begin{gathered}
 \check{\varepsilon}\backslash\check{\mu} = \varepsilon^p\backslash{\mu^p}
 \begin{pmatrix}
  \frac{g(t)}{h(t,y)} & 0 & 0\\
  0 & \frac{1}{g(t)} & 0\\
  0 & 0 & \frac{g(t)}{h(t,y)}
 \end{pmatrix}, 
\\
\check{\gamma_1} = \check{\gamma_2}^{\mathtt{T}} = \frac{yg(t)\dot{g}(t)\varepsilon^p\mu^p}{h(t,y)}
 \begin{pmatrix}
  0 & 0 & 1 \\
  0 & 0 & 0 \\
  \text{-}1 & 0 & 0
 \end{pmatrix},
\end{gathered}
\end{equation}
where we use the notation $\check{\varepsilon}\backslash\check{\mu}$ to denote either $\check{\varepsilon}$ or $\check{\mu}$, an overdot $\dot{g}(t)$ denotes a time derivative, and
\begin{equation}
 h(t,y)=1-y^2\dot{g}^2(t)\varepsilon^p\mu^p.
\end{equation}

In this case the resulting material parameters are not so trivial as the previous examples, depending on a combination of \textit{both} the prior dielectric material parameters $\varepsilon^{p}$ and ${\mu^p}$. It is not obvious how such a result could be anticipated from the Plebanski-De Felice approach, which is formulated specifically for vacuum space-times. In the following examples it will be clear that a naive application of the Plebanski-De Felice approach fails for more complicated prior dielectrics. 

\section{Anisotropic Prior Dielectric} \label{Sec:Anisotropic}
It is interesting that effects such as those seen in the last example arise even at the level of an isotropic prior dielectric.  The calculations can just as easily be repeated for an arbitrary anisotropic prior, where the results can become significantly more complicated.  For simplicity, consider a prior anisotropic dielectric material described by $\check{\varepsilon}^p = diag(\varepsilon_{xx}^p,\varepsilon_{yy}^p,\varepsilon_{zz}^p)$ and $\check{\mu}^p = diag(\mu_{xx}^p,\mu_{yy}^p,\mu_{zz}^p)$, with vanishing magnetoelectric coupling.  

Repeating the calculations for the transformation of Eq.\ (\ref{Eq:TimeTrans}) now leads to the material parameters
\begin{equation} \label{Eq:AnisoSTtrans}
\begin{gathered}
 \check{\varepsilon} = \frac{1}{f(x)} 
 \begin{pmatrix}
  \varepsilon_{xx}^p&0&0\\
  0&\varepsilon_{yy}^p&0\\
  0&0&\varepsilon_{zz}^p
 \end{pmatrix}, \\
 \check{\mu}= \frac{1}{f(x)}
 \begin{pmatrix}
  \mu_{xx}^p&0&0\\
  0&\mu_{yy}^p&0\\
  0&0&\mu_{zz}^p
 \end{pmatrix},\\
\check{\gamma_1} = \check{\gamma_2}^{\mathtt{T}} = \frac{f'(x)t}{f(x)}
 \begin{pmatrix}
  0 & 0 & 0 \\
  0 & 0 & 1 \\
  0 & \text{-}1 & 0
 \end{pmatrix}.
\end{gathered}
\end{equation}
This result may have been anticipated based on the results found in Eq.\ (\ref{Eq:IsoSTtrans}).  Do the results of the time dependent spatial transformation in a prior isotropic dielectric, Eq.\ (\ref{Eq:IsoTStrans}), allow us to anticipate the result of the same time dependent spatial transformation in a prior anisotropic dielectric?  Repeating the calculations for the transformation of Eq.\ (\ref{Eq:SpaceTrans}) with a prior anisotropic dielectric leads to the material parameters
\begin{equation}
  \begin{gathered}
 \check{\varepsilon} =
 \begin{pmatrix}
  \frac{g(t)\varepsilon_{xx}^p}{1-y^2\dot{g}^2(t)\varepsilon_{xx}^p\mu_{zz}^p}&0&0\\
  0&\frac{\varepsilon_{yy}^p}{g(t)}&0\\
  0&0&\frac{g(t)\varepsilon_{zz}^p}{1-y^2\dot{g}^2(t)\varepsilon_{zz}^p\mu_{xx}^p}
 \end{pmatrix},\\
 \check{\mu}= 
 \begin{pmatrix}
  \frac{g(t)\mu_{xx}^p}{1-y^2\dot{g}^2(t)\varepsilon_{zz}^p\mu_{xx}^p}&0&0\\
  0&\frac{\mu_{yy}^p}{g(t)}&0\\
  0&0&\frac{g(t)\mu_{zz}^p}{1-y^2\dot{g}^2(t)\varepsilon_{xx}^p\mu_{zz}^p}
 \end{pmatrix},\\
\check{\gamma_1} = yg(t)\dot{g}(t)
 \begin{pmatrix}
  0 & 0 & \frac{\varepsilon_{zz}^p\mu_{xx}^p}{1-y^2\dot{g}^2(t)\varepsilon_{zz}^p\mu_{xx}^p} \\
  0 & 0 & 0 \\
  \frac{\text{-}\varepsilon_{xx}^p\mu_{zz}^p}{1-y^2\dot{g}^2(t)\varepsilon_{xx}^p\mu_{zz}^p} &  & 0
 \end{pmatrix},
\end{gathered}
\end{equation}
with $\check{\gamma_2}=\check{\gamma_1}^{\mathtt{T}}$.  Indeed, this result mirrors several features of Eq.\ (\ref{Eq:IsoTStrans}), but the additional complexity of the anisotropic dielectric contributes some slight differences between the two sets of results.  Carefully note, for example, how the $xx$ and $zz$ components of $\check{\varepsilon}$ and $\check{\mu}$ depend on the combinations $\varepsilon_{zz}\mu_{xx}$ and $\varepsilon_{xx}\mu_{zz}$, and similarly for the components of $\gamma_1$ and $\gamma_2$.

\subsection{Square Cloak} \label{Sec:SquareCloak}
The transformations considered thus far have been relatively simple and generic.  Consider the more concrete and well-known example of a square cloak \cite{Rahm:200887}, embedded in the prior anisotropic dielectric described previously.  In this case the transformation is
\begin{equation}
 \mathcal{T}(t,x,y,z)=\left(t,\frac{s_2(x-s_1)}{(s_2-s_1)},\frac{ys_2(x-s_1)}{x(s_2-s_1)},z\right),
\end{equation}
for constants $s_1$ and $s_2$, resulting in the material parameters
\begin{equation}
 \check{\varepsilon}=
 \begin{pmatrix}
  \frac{(x-s_1)\varepsilon^p_{xx}}{x} & * & *\\
  -\frac{y s_1 \varepsilon^p_{xx}}{x^2} & \frac{y^2s_1^2\varepsilon^p_{xx}+x^4\varepsilon^p_{yy}}{x^3(x-s_1)} & *\\
  0 & 0 & \frac{s_2^2(x-s_1)\varepsilon^p_{zz}}{x(s_2-s_1)^2}
 \end{pmatrix},
\end{equation}
and similarly for $\check{\mu}$, while $\check{\gamma_1}=\check{\gamma_2}=0$.  This should be compared to the vacuum result obtained by setting $\varepsilon^p_{xx}=\varepsilon^p_{yy}=\varepsilon^p_{zz}=1$. Because the square cloak transformation is purely spatial and the prior dielectric has no magnetoelectric coupling, this result should in principle be obtainable with the approach of Ref.\ \cite{Milton:2006}, based on an explicit application of Maxwell's equations.  However, it is clearly not obtainable from the Plebanski-De Felice equations.

\section{Magnetoelectric Prior Dielectric} \label{Sec:MagnetoElectric}
So far the discussion has focused on prior dielectrics with vanishing magnetoelectric coupling.  The existing literature gives no indication as to how a prior dielectric with non-vanishing magnetoelectric couplings should be accommodated, but the completely covariant approach facilitates such a prior material just as easily as any other prior material and one proceeds in exactly the same way in all cases.  Let a prior dielectric material with non-vanishing magnetoelectric couplings be described by scalar-valued $\varepsilon^p$ and $\mu^p$, and magnetoelectric couplings
\begin{equation}
 \check{\gamma_1}^p=(\check{\gamma_2}^p)^{\mathtt{T}} =
 \begin{pmatrix}
  0 & \gamma^p_{xy} & \gamma^p_{xz}\\
  \text{-}\gamma^p_{xy} & 0 & \gamma^p_{yz}\\
  \text{-}\gamma^p_{xz} & \text{-}\gamma^p_{yz} & 0
 \end{pmatrix}.
 \end{equation}
These material parameters are given in the three-dimensional representation of Eq.\ (\ref{Eq:ConstitutiveComponents2}), but the calculations proceed with $\boldsymbol{\chi}$, which is more intimately related to the three-dimensional representation of Eq.\ (\ref{Eq:ConstitutiveComponents1}).  So it is important to first determine the parameters of $\boldsymbol{\chi}$ by changing to the representation of Eq.\ (\ref{Eq:ConstitutiveComponents1}).  We are now in a position to reconsider the transformations given by Eqs. (\ref{Eq:YInversion})--(\ref{Eq:TimeTrans}), and (\ref{Eq:SpaceTrans}).

The spatial scaling of Eq.\ (\ref{Eq:YInversion}) now leads to $\check{\varepsilon}=\text{-}\varepsilon^p diag(a^{\text{-}1},a,a^{\text{-}1})$, $\check{\mu}=\text{-}\mu^p diag(a^{\text{-}1},a,a^{\text{-}1})$, and
\begin{equation}
 \check{\gamma_1}=\check{\gamma_2}^{\mathtt{T}} =
 \begin{pmatrix}
  0 & \gamma^p_{xy} & \text{-}\frac{\gamma^p_{xz}}{a}\\
  \text{-}\gamma^p_{xy} & 0 & \gamma^p_{yz}\\
  \frac{\gamma^p_{xz}}{a} & \text{-}\gamma^p_{yz} & 0
 \end{pmatrix}.
\end{equation}
Note that if this magnetoelectric coupling can be interpreted as a velocity \cite{Leonhardt:2006a} (which is not always possible \cite{Thompson:2010a}), then we must rescale its $y$-component relative to such a velocity in the vacuum case.  On the other hand the temporal scaling of Eq.\ (\ref{Eq:TInversion}) now leads to $\varepsilon = \text{-}a\varepsilon^p$, $\mu=\text{-}a\mu^p$, $\check{\gamma_1}=\text{-}a\check{\gamma_1}^p$, and $\check{\gamma_2}=\text{-}a\check{\gamma_2}^p$.

The spatially dependent time transformation of Eq.\ (\ref{Eq:TimeTrans}) now leads to $\varepsilon = \varepsilon^p/f(x)$, $\mu = \mu^p/f(x)$, and
\begin{equation}
 \check{\gamma_1}=\check{\gamma_2}^{\mathtt{T}} = \frac{1}{f(x)}
 \begin{pmatrix}
    0 & \gamma^p_{xy} & \gamma^p_{xz}\\
  \text{-}\gamma^p_{xy} & 0 & \gamma^p_{yz}+tf'(x)\\
 \text{-}\gamma^p_{xz} & \text{-}\gamma^p_{yz}-tf'(x) & 0
 \end{pmatrix}.
\end{equation}

Lastly, the time dependent spatial transformation of Eq.\ (\ref{Eq:SpaceTrans}) now leads to
\begin{equation}
\begin{gathered}
  \check{\varepsilon}\backslash\check{\mu} = \frac{\varepsilon^p\backslash\mu^p}{w}
  \begin{pmatrix}
   g(t) & y\dot{g}(t)\gamma^p_{yz} & 0\\
   * & \frac{w+y^2\dot{g}^2(t)\left((\gamma^p_{xy})^2+(\gamma^p_{yz})^2\right)}{g(t)} & y\dot{g}(t)\gamma^p_{xy} \\
  * & * & 0
  \end{pmatrix}, \\
  \check{\gamma_1}=\check{\gamma_2}^{\mathtt{T}} = \frac{1}{w}
  \begin{pmatrix}
   0 & z\gamma^p_{xy} & g(t)\left(y\dot{g}(t)\varepsilon^p\mu^p +z\gamma^p_{xz}\right) \\
  * & 0 & z\gamma^p_{yz}\\
  * & * & 0.
  \end{pmatrix},
\end{gathered}
\end{equation}
where $*$ indicates a matrix element whose value is obtained by the appropriate symmetry properties of the matrix ($\check{\varepsilon}$ and $\check{\mu}$ are symmetric while $\check{\gamma_1}$ and $\check{\gamma_2}$ are antisymmetric)
\begin{equation}
 z = (1-y\dot{g}(t)\gamma^p_{xz}),\  \text{and}\  w = z^2-y^2\dot{g}^2(t)\varepsilon^p\mu^p.
\end{equation}
This example demonstrates a case where neither the Plebanski-De Felice approach nor the approach of Ref.\ \cite{Milton:2006} are applicable, but is handled easily with the totally covariant approach.  The result clearly illustrates the complexity that may arise, even for simple transformations, when the prior dielectric material is complicated, and in general a particular component of the final material may depend on several parameters of the prior material.

\section{Discussion} \label{Sec:Conclusions}
Using a totally covariant approach, I have demonstrated the extension of transformation optics to nonvacuum scenarios.  I have shown that for transformation optics in prior nonvacuum dielectric media, the material parameters of the prior media may appear in the final material parameters in complicated and unpredictable ways.  The exact manner in which the prior material parameters appear naturally depends both on the characteristics of the prior material (i.e.\ whether it is isotropic or anisotropic, and whether there are non-vanishing magnetoelectric couplings), and on the transformation.  Strictly speaking it is not possible to obtain such results from the Plebanski-De Felice equations, as these are formulated for vacuum space-times.  Because of the relatively trivial nature of an isotropic material, using the Plebanski-De Felice equations and a naive scaling of the vacuum dielectric parameters with the material parameters happens to work for simple transformations.  For more complicated materials and transformations this naive approach no longer works and the completely covariant approach becomes necessary.

When the transformation is a spatially dependent scaling of time such as Eq.\ (\ref{Eq:TimeTrans}), the required material parameters are a simple scaling of the prior material parameters.  On the other hand, time dependent spatial transformations such as that of Eq.\ (\ref{Eq:SpaceTrans}) generally appear to require a material whose parameter complexity increases with the complexity of the prior dielectric.  An exotic time dependent transformation is not required to see interesting and unanticipated behaviors, as the familiar square cloak embedded in an anisotropic dielectric demonstrates.

Although the example transformations and initial dielectric materials reported here were necessarily limited, I wish to emphasize that the completely covariant formulation of transformation optics encompasses all types of transformations and arbitrarily complex linear dielectric materials, and that the calculations proceed identically in all cases.  As transformation optics designs become more complex, the completely covariant approach will be needed to handle the required transformations and materials.

\begin{acknowledgments}
 I thank Steve Cummer and J\"{o}rg Frauendiener for useful comments and discussions.
\end{acknowledgments}

\end{document}